# Revealing the configurations and host-guest interactions of small aromatics confined in porous frameworks by electron microscopy

Authors: Boyuan Shen[1], Xiao Chen[1*], Hao Xiong[1], Fei Wei[1*]

Affiliations:
[1] Beijing Key Laboratory of Green Chemical Reaction Engineering and Technology, Department of Chemical Engineering, Tsinghua University, Beijing 100084, China.
*Corresponding author:
Email: chenx123@tsinghua.edu.cn (X.C.); wf-dce@tsinghua.edu.cn (F.W.)

**Abstract**
Directly imaging the configurations of small molecules at the ambient temperatures will greatly promote the study of their chemical and physical properties, including the host-guest interactions of organics in porous materials during the adsorption, catalysis and energy storage. However, due to the current challenges on the small-molecule imaging by the (scanning) transmission electron microscopy ((S)TEM), we still have a lack of the molecular-level understandings on the host-guest interactions and other molecular behaviors. Here, we achieved the STEM imaging of various small aromatics confined in the MFI-type zeolite frameworks by using the integrated differential phase contrast (iDPC) technique. Due to the strong confinement effect in MFI channels, the 1D solid-like aromatic columns showed the coherent configurations, which were clearly resolved by enhancing the host-guest interactions. Then, we also evaluated the strength of host-guest interactions directly by the image analysis and revealed the desorption behaviors of confined aromatics during the *in-situ* heating process. These results not only helped us to reveal the configurations and host-guest interactions of small aromatics during the adsorption/desorption in porous materials, but also expanded the applications of STEM to further study other molecular behaviors in the real space.

**Introduction**
Achieving small-molecule imaging under diverse microscopes is always a milestone in nanotechnology and molecular science. The real-space observations of small molecules are of great significance for understanding their chemical and physical properties[1-6]. For example, directly resolving the molecular configurations in the porous frameworks helps us visually investigate the host-guest interactions during the adsorption, catalysis, gas storage and separation[7-13], which are the key issues in these applications but poorly understood at the molecular level. When the sizes of the guest molecules and the host channels are comparable, the configurations of guest molecules will be stabilized by reciprocally optimizing the host-guest interactions, known as the confinement effect[14,15]. Such effect will reduce the rapid switches between the positions and configurations of small molecules, which can be further revealed by the real-space imaging.
The (scanning) transmission electron microscopes ((S)TEMs) are the most powerful imaging tools to obtain the structure information with the atomic resolution[16-19], which is obviously enough to detect the small molecules in the porous frameworks. However, the typical porous materials in these applications and the weak host-guest bonding are easily damaged by high-energy electron beams[20-24]. And the ultra-low contrasts of light elements in organic molecules are usually submerged in the imaging noises. The recent developments of a low-dose imaging technique, called the integrated differential phase contrast (iDPC) STEM (Fig. S1 and Text S1)[19,25-27], showed its potential advantages for imaging the beam-sensitive materials and the light-element components. This technique provides us the possibility to observe the small organic molecules in porous frameworks under the STEM.

Here, we confined various aromatic molecules in the MFI-type zeolite frameworks by gradually enhancing the interactions. These aromatics underwent a transition from the liquid phase into a 1D solid-like phase of aromatic columns with coherent orientations due to the strong confinement effect. Then, we used the iDPC-STEM imaging to reveal the overlapped configurations (orientations) of these aromatic columns. Based on the relation between the characteristics of aromatic images and the strength of interactions, we proposed a new method to estimate the host-guest interactions between different aromatics and frameworks by the image analysis. These results established a general method to image the small molecules by electron microscopy after confined in the size-matched channels, which advanced our understandings on the confinement effect and adsorption/desorption behaviors of small molecules in porous frameworks.

**The strategy to image aromatic configurations by enhanced interactions**

Fig. 1 exhibits the schematics of our strategy to confine and image small aromatics by enhancing the host-guest interactions. When adsorbed by a size-matched channel, the small aromatics will be frozen to form a 1D solid-like phase (an aromatic column with nearly coherent orientation) by the strong confinement effect. For example, we confined the benzene molecules in the straight channels of a MFI-type Si/O framework (Fig. S2 and S3), named the silicalite-1[28]. And we can directly image these benzene columns in channels by the iDPC-STEM, which is operated at 300 kV with an electron dose less than 40 $e^-/Å^2$. Then, we changed the benzenes to the pyridines that show a similar kinetic diameter, and also changed the silicalite-1 to the ZSM-5[29], another Al-doping MFI-type framework with nearly the same framework structure. At the acid site in zeolites, a bridged hydroxyl group is formed by a proton attached to the O atom in the Al-O-Si unit. And, this proton will be accepted by the confined base species, such as the pyridine, to form the protonated base $BH^+$. The $BH^+$ is stabilized by the electrostatic interaction with the deprotonated zeolite $Z^-$. The net negative charge in the $Z^-$ is easily spread out and provides a diffuse electron cloud over the inner surface of framework, indicating that the $Z^-$ can be considered as a soft conjugated base to stabilize the whole protonated pyridine columns confined in channels[30-32]. Moreover, the $Si_{10}$-rings in ZSM-5 channels will deform to show a less circular shape than that of silicalite-1 channels (more detailed discussions in following text). The size and shape matching between guest and host will enhance both the van der Waals and electrostatic interactions. Thus, the pyridines can be imaged with more stable configurations and higher contrasts in the ZSM-5 channels by the iDPC-STEM.

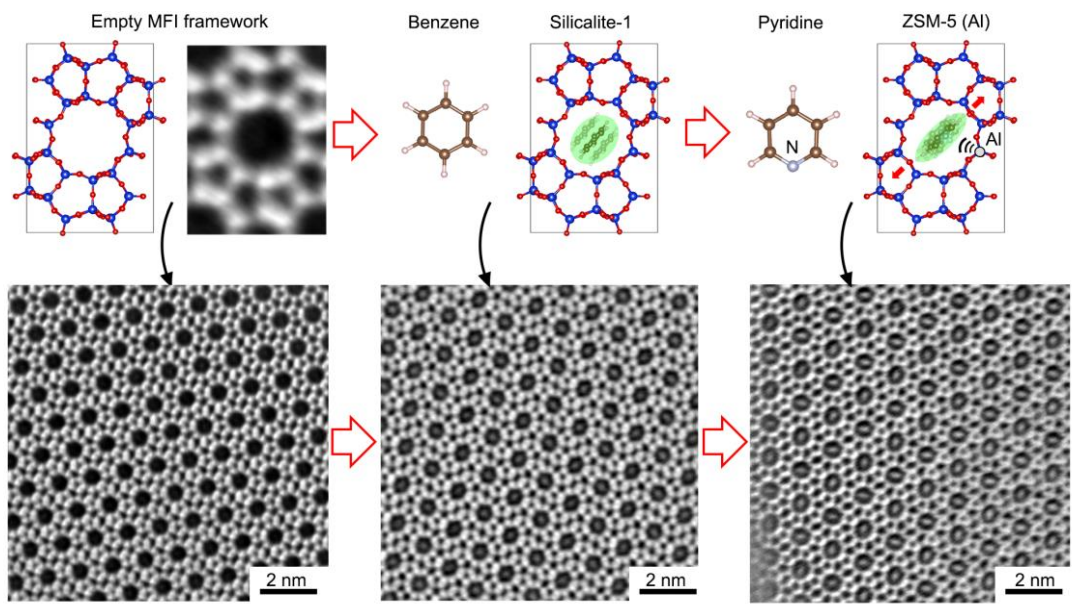

**Fig. 1 | The schematics of the strategy to freeze and resolve the small aromatics by enhancing the interactions and confinements.** We imaged the benzenes confined in the silicalite-1 and the pyridines confined in the ZSM-5. The enhanced confinement in the pyridine/ZSM-5 specimen, due to the additional electrostatic interaction and the slightly changed channel geometry, resulted in a clearer image of pyridine columns with sharper shapes and higher contrasts.

**Imaging results of the benzenes and pyridines confined in zeolites**

Fig. 2 provides the imaging and analysis of the aromatic columns confined in two MFI-type frameworks. Before filling the MFI-type frameworks (the silicalite-1 and ZSM-5) with benzenes and pyridines, we can clearly observe the Si (or Al) atoms on the [010] projection without any contrast (intensity) in the empty straight channels as shown in Fig. 1 and S4. And, after the aromatic adsorption, some visible species appeared in these empty channels. For example, Fig. 2a provides the iDPC image of the benzene columns confined in the straight channels of silicalite-1. The magnified image in Fig. 2b shows an obvious elliptical spot inside the single channel with its long axis nearly parallel to that of the elliptical $Si_{10}$-ring. Due to the strong host-guest *van der Waals* interactions, the straight channels of the silicalite-1 (with the size of ~5.6×5.3 Å$^2$) show the efficient confinement effect to various monocylic aromatics (with the kinetic diameter of ~5.8 Å). Theoretically, the $C_6$-planes in these monocylic aromatics preferred to be parallel to the b-axis of MFI-type framework energetically, and oriented along the long axis of $Si_{10}$-rings[32-37]. Thus, the liquid aromatics were frozen into a 1D solid-like phase by the strong confinement effect at the ambient temperature, although they would still vibrate near the equilibrium positions within the dwell time of scanning. Then, the integrated contrast of the whole aromatic column could be detected as an elliptical spot with the molecular resolution.

To better describe and compare the imaged aromatic spots, we used the profile analysis of iDPC-STEM images to define the spot sizes. The profile directions were nearly along the long and short axes of aromatic spots as marked by the red and blue arrows in Fig.

2b respectively. As measured in Fig. 2c, the peaks in the intensity profiles of the filled channels indicated the benzene spot with a size of 3.65×3.90 Å$^2$ compared to those of empty channels (Fig. S5). Then, we used the ratio of peak widths from two directions (3.90/3.65, defined as the aspect ratio) to roughly describe the dynamic information of aromatic columns during the electron probe scanning and estimate the strength of host-guest interactions.

On the basis of the benzenes/silicalite-1 system, where we only introduced the *van der Waals* host-guest interactions, we also established the pyridine/ZSM-5 system with very similar structures but enhanced host-guest interactions. The iDPC-STEM image in Fig. 2d provides the imaged pyridine columns in the straight channels of ZSM-5 framework. Compared to the benzene/silicalite-1 system, the pyridine spot obviously owns a higher contrast, sharper elliptical shape and clearer orientation resulting from the additional electrostatic interactions between the protonated ZSM-5 and the protonated pyridines. In the magnified image and corresponding profile analysis (Fig. 2e and f), we observed that the pyridine columns in ZSM-5 showed a higher aspect ratio (sharper shape) than that of the benzene spot in silicalite-1. In order to further explain the relation between the aromatic images and the host-guest interactions, we also provided the images of the pyridine/silicalite-1 and benzene/ZSM-5 specimens as comparisons in Fig. 2g-l. In these two systems, the aromatics were also confined in channels as the 1D solid-like phase mainly by *van der Waals* interactions and, thus, showed the similar shapes and orientations with the benzenes in silicalite-1.

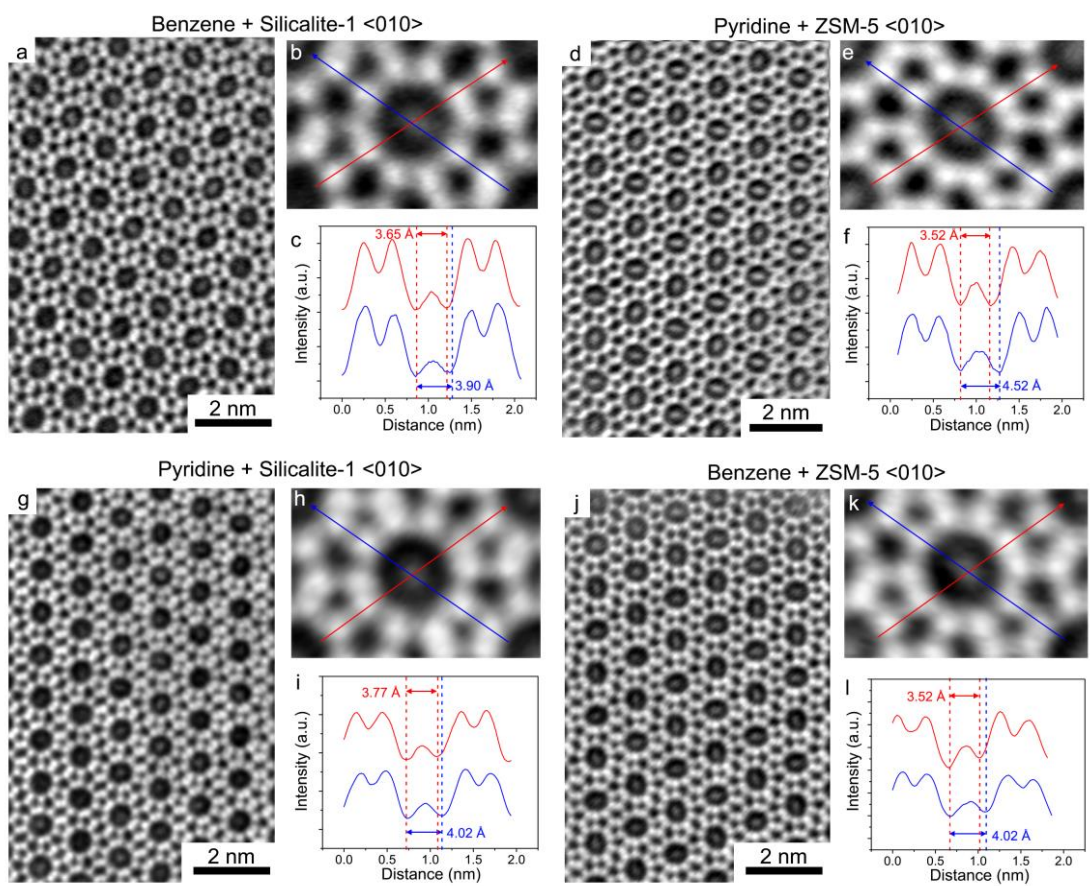

**Fig. 2 | Imaging the aromatic columns confined in the MFI-type frameworks. a-c**, The iDPC-STEM images and profile analysis of the benzene/silicalite-1 specimen from the <010> projection. The magnified image shows a clear aromatic spot observed inside the straight channel. And the profile analysis gives the aspect ratio of the aromatic spot to estimate the host-guest interaction. **d-f**, The iDPC-STEM images and profile analysis of the pyridine/ZSM-5 specimen. **g-i**, The iDPC-STEM images and profile analysis of the pyridine/silicalite-1 specimen. **j-l**, The iDPC-STEM images and profile analysis of the benzene/ZSM-5 specimen.

**Evaluating the host-guest interactions by image analysis**

Furthermore, we analyzed the impact of the strength of host-guest interaction strength on the imaging results based on these four specimens. As we mentioned above, in the size-matched channels, there is a transition from the bulk phase of aromatics (liquid or gas) into a 1D solid-like phase, where these aromatic columns were confined in these channels with the coherent orientations (as shown in Fig. 3a). The strong host-guest interactions, including the van der Waals and the electrostatic interactions, will help to enhance the confinement effect to stabilize the aromatic configurations in a deep energy trap. Reflected in the images, the aromatic columns show the sharper shapes (higher aspect ratios) and higher contrasts. While, the lower interactions will induce the switch between configurations with similar energies, the molecular vibration near equilibrium positions or even the aromatic adsorption from channels. Reflected in the images, the aromatic columns show the near round shapes (lower aspect ratios) and lower contrasts. In Fig. 3b, we measured the average aspect ratios of aromatics in these four specimens imaged by the iDPC-STEM. As we expected, the benzene and pyridine columns in the silicalite-1 were imaged with nearly the same aspect ratios, since they own very similar molecular sizes and were confined in the same framework. And, we found the aspect ratios of benzene columns increased a little in ZSM-5, which is mainly attributed to the slightly changed channel geometry after Al doping in ZSM-5. To confirm, we measured the projected distances between the opposite Si atoms along the same intensity profiles in Fig. 2 (details in Fig. S6), and used the ratios of these two atom distances to identify the aspect ratios of channels. As shown in Fig. 3c, the average aspect ratios of channels in the ZSM-5 are obviously higher than those in the silicalite-1. Both the van der Waals and electrostatic interactions are dependent on how the sizes and shapes of aromatics fit with those of zeolite channels. Thus, the higher aspect ratios of ZSM-5 channels will stabilize the imaged aromatic configuration that is parallel to the long axis of elliptical channel, resulting in the ranking of aspect ratios in Fig. 3b.

Then, we could also use the *in-situ* observations of aromatic desorption behaviors to reveal the host-guest interactions. In Fig. 3d, we found the shapes and orientations of pyridines in ZSM-5 could also be clearly imaged at 250 °C during the *in situ* heating process. At 350 and 450 °C, the contrasts in channels were reduced due to the gradually desorbed pyridines, but we could still observe some low-contrast species retained in channels, which may represent the strong binding of pyridines at the Al acid sites. Then, after heated to 500 °C, all the pyridines were desorbed from the ZSM-5, and the straight channels became empty. And, after we cooled the specimen down to 100 °C, the straight channels were still empty without any species inside (details in Fig. S7). On the one hand, these results showed that such strategy to freeze and image small molecules still worked at higher temperatures until they were completely desorbed. On the other hand, these results could help us to investigate the *in-situ* adsorption behaviors of various aromatics in porous frameworks at the molecular level.

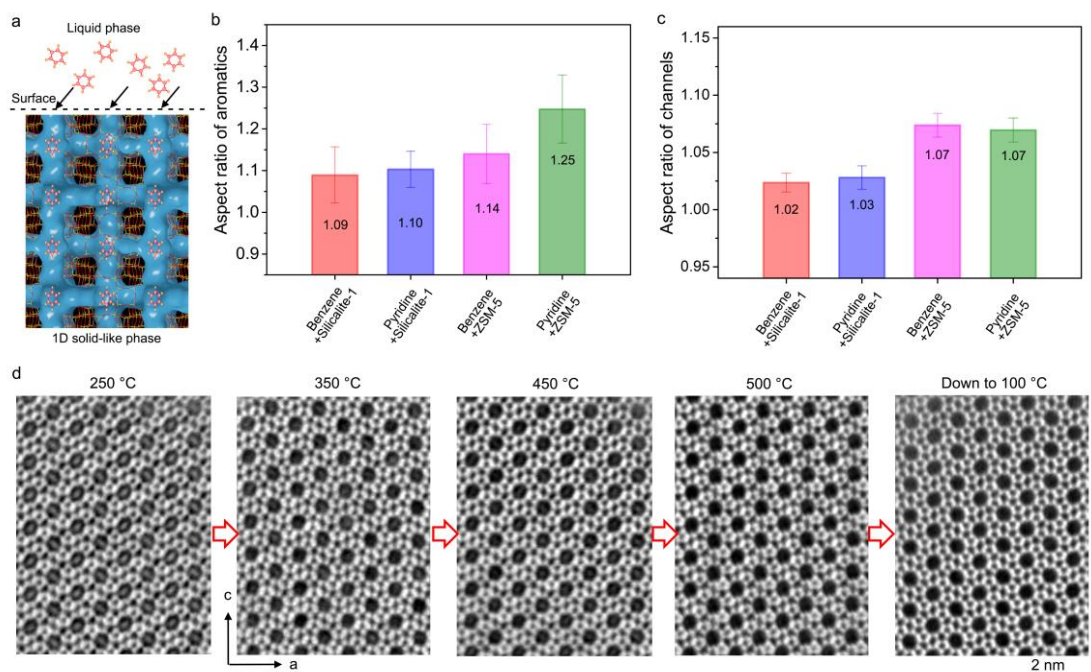

**Fig. 3 | The host-guest interaction studied by the image analysis. a**, The schematics showing the transition of aromatics from the liquid phase to the 1D solid-like phase (the aromatic columns in the MFI-type frameworks). **b**, The statistics of the aspect ratios of aromatic columns in four specimens, which indicate the different strength of host-guest interactions. The error bars represent the standard deviations of 20 data for each specimen. **c**, The statistics of the aspect ratios of channels in four specimens, where the ZSM-5 channels show the higher aspect ratios than those of the silicalite-1 channels. The error bars represent the standard deviations of 20 data for each specimen. **d**, The *in-situ* iDPC-STEM images of the pyridine/ZSM-5 specimen at different temperatures (from 250°C to 500 °C then back to 100 °C).

**More discussions and evidences for heterocyclic aromatics**

Then, we applied this imaging method to more confined-aromatic systems, and estimate the host-guest interactions in these systems by the analysis of images. Fig. 4 shows the images of three heterocyclic aromatics in ZSM-5 under the same specimen preparation conditions. The thiophenes, pyrroles and furans are all heterocyclic aromatics with five membered rings, but their molecular size and polarity are a little different. As we have observed, the ranking of the contrasts and aspect ratios of these aromatics is thiophene > pyrrole > furan, which is also confirmed by the statistics in Fig. 4j. And based on our discussions above, the ranking of aspect ratios is a reflection of the comparison between interaction strengths in the images. The confinement effect in zeolites is dependent on both the long-range van der Waals and electrostatic host-guest interactions, which also influenced by many factors of guest molecules, such as the size, polarity and basicity

(proton affinity). And, the real-space observations provided a comprehensive outcome of the complex interactions, which can be used to estimate the collaboration of various factors.

In these systems, the thiophene owns a similar molecular size with those six-membered aromatics, which is obviously larger than that of pyrrole and furan since the S atom has one more electronic shell than N and O atoms in pyrrole and furan respectively. Thus, the matched size and polarity provided the higher aspect ratio for the thiophenes in spite of only van der Waals interactions included. As for the pyrrole, although it owns a N basic site like the pyridine, the lone-pair electrons on the pyrrole N atom will participate in the formation of conjugated π bonds, which results in its much lower proton affinity than that of the pyridine. Additional electrostatic interaction in pyrrole/ZSM-5 system is too weak to fill in the difference in van der Waals interactions between pyrrole/ZSM-5 and thiophene/ZSM-5 systems. Thus, the aspect ratio of pyrroles in ZSM-5 is lower than that of thiophenes. And, naturally, the small-size furan without basicity shows the lowest aspect ratio and host-guest interaction.

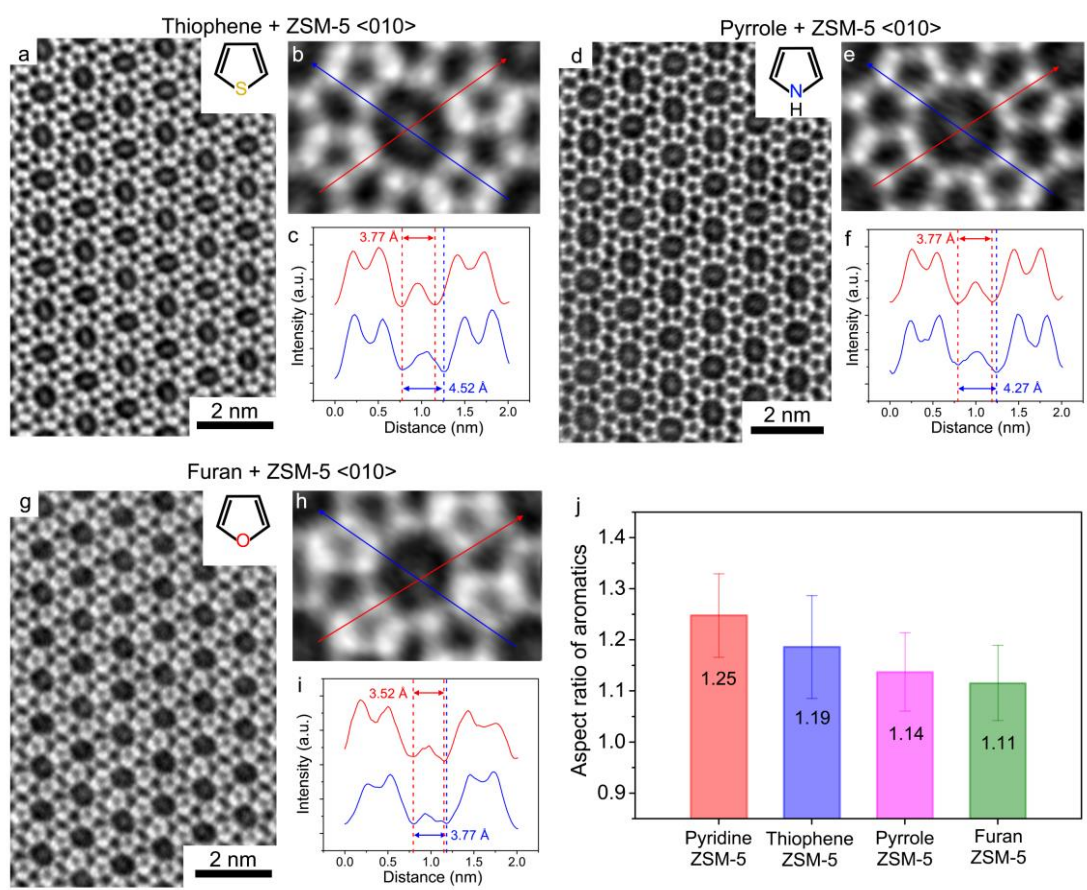

**Fig. 4 | Imaging other heterocyclic aromatic columns confined in the ZSM-5 frameworks. a-c**, The iDPC-STEM images and profile analysis of the thiophenes in the ZSM-5 framework from the <010> projection. **d-f**, The iDPC-STEM images and profile analysis of the pyrroles in the ZSM-5 framework. **g-i**, The iDPC-STEM images and profile analysis of the furans in the ZSM-5 framework. **j**, The statistics of the aspect ratios of four aromatic columns in ZSM-5 framework, which indicate the different strength of host-guest interactions. The error bars represent the standard deviations of

20 data for each specimen.

In summary, based on the low-dose iDPC-STEM technique, we achieved the real-space imaging of various small aromatics adsorbed in zeolite frameworks. The configurations of aromatic columns were stabilized in the straight channels by the strong confinement effect, which can be directly revealed by the iDPC-STEM imaging. And, we established a new method to investigate the host-guest interactions and the adsorption/desorption behaviors of small molecules by the image analysis, which expanded the applications of electron microscopy for the molecular-scale characterization in catalysis, gas storage and separation. Meanwhile, these results also indicated a probably general strategy to freeze the configurations of other complex molecules by the sized-matched channels, and resolve them at the ambient or even higher temperatures. Thus, more molecules and their behaviors can be imaged and studied during the *in situ* experiments, where their configurations are closer to those at the real conditions in different applications.

**Materials and Methods**
The synthesis of MFI-type zeolites
The short-b-axis MFI-type zeolite crystals, including the ZSM-5 and silicalite-1, were synthesized by the hydrothermal method. For the ZSM-5, the tetrapropylammonium hydroxide (TPAOH, 25% in water, 13.1 g), tetraethylorthosilicate (TEOS, 11.2 g), iso-propanol (IPA, 0.1 g), urea (2.0 g), $Al(NO_3)_3·9H_2O$ (0.3g) and NaOH (0.1 g) were used as reactants and stirred in 18.4 g $H_2O$ for 2 h. Then the mixture was transferred into an autoclave for the further crystallization. Before the mixture was quenched by cold water, it was heated to 180 °C with a rate of 15 °C/h and kept at 180 °C for another 48 h. The deionized water was used to wash the as-prepared Na-ZSM-5 crystals. The crystals were calcined at 500 °C for over 5 h to remove the templates. Then, the $NH_4$-ZSM-5 crystals were obtained by the cation exchange in a 1 mol/L $NH_4NO_3$ solution. Finally, the H-ZSM-5 crystals were obtained after calcining the NH4-ZSM-5 at 550 °C for over 5 h. And the silicalite-1 crystals were synthesized by nearly the same process without adding the $Al(NO_3)_3·9H_2O$ and NaOH.

Filling the MFI-type zeolites with aromatics
The MFI-type zeolite frameworks were filled by the aromatics, including the benzenes, pyridines, thiophenes, pyrroles and furans, directly in the pure aromatic liquid under the ultrasound for 30 min. And the MFI-type zeolites were kept in the liquid for a long time for the sufficient diffusion of aromatics. Before the iDPC-STEM imaging, the zeolite crystals were fully dispersed in the liquids by the ultrasound.

The iDPC-STEM imaging
A Cs-corrected STEM (FEI Titan Cubed Themis G2 300) was operated at 300 kV to obtained the iDPC-STEM images. The STEM was equipped with a DCOR+ spherical aberration corrector for the electron probe which was aligned using a standard gold sample before observations. The aberration coefficients are C1=2.27 nm; A1=0.41 nm; A2=1.76 nm; B2=15.1 nm; C3=217 nm; A3=417 nm; S3=97.6 nm; A4=2.16 μm,

D4=3.77 μm, B4=6.69 μm, C5=187 μm, A5=168 μm, S5=55.2 μm, and R5=13.4 μm. The convergence semi-angle is 10 mrad. The collection semi-angles of iDPC-STEM images are 4-20 mrad. The beam current is lower than 0.1 pA, corresponding to a low electron dose of 40 e$^-$/Å$^2$. The *in situ* heating experiment was conducted using the in situ holder (Protochips, Fusion 350), which was heated to different target temperatures with an ultrahigh heating rate of 1000 °C/ms.

**References**
1. H. J. Lee, W. Ho, Single-bond formation and characterization with a scanning tunneling microscope. *Science* **286**, 1719-1722 (1999).
2. R. Temirov, S. Soubatch, A. Luican, F. S. Tautz, Free-electron-like dispersion in an organic monolayer film on a metal substrate. *Nature* **444**, 350-353 (2006).
3. L. Gross, F. Mohn, N. Moll, P. Liljeroth, G. Meyer, The chemical structure of a molecule resolved by atomic force microscopy. *Science* **325**, 1110-1114 (2009).
4. L. Gross *et al.*, Organic structure determination using atomic-resolution scanning probe microscopy. *Nat. Chem.* **2**, 821-825 (2010).
5. C. Weiss, C. Wagner, R. Temirov, F. S. Tautz, Direct imaging of intermolecular bonds in scanning tunneling microscopy. *J. Am. Chem. Soc.* **132**, 11864-11865 (2010).
6. J. Zhang *et al.*, Real-space identification of intermolecular bonding with atomic force microscopy. *Science* **342**, 611-614 (2013).
7. M. Eddaoudi et al., Systematic design of pore size and functionality in isoreticular MOFs and their application in methane storage. *Science* **295**, 469-472 (2002).
8. Z. Lai et al., Microstructural optimization of a zeolite membrane for organic vapor separation. *Science* **300**, 456-460 (2003).
9. N. L. Rosi et al., Hydrogen storage in microporous metal-organic frameworks. *Science* **300**, 1127-1129 (2003).
10. J. Lee et al., Metal-organic framework materials as catalysts. *Chem. Soc. Rev.* **38**, 1450-1459 (2009).
11. J. R. Li, R.J. Kuppler, H. C. Zhou, Selective gas adsorption and separation in metal–organic frameworks. *Chem. Soc. Rev.* **38**, 1477-1504 (2009).
12. M. Choi et al., Stable single-unit-cell nanosheets of zeolite MFI as active and long-lived catalysts. *Nature* **461**, 246-249 (2009).
13. P. J. Bereciartua et al., Control of zeolite framework flexibility and pore topology for separation of ethane and ethylene. *Science* **358**, 1068-1071 (2017).
14. E. G. Derouane, J. M. Andre, A. A. Lucas, Surface Curvature Effects in Physisorption and Catalysis by Microporous Solids and Molecular Sieves. *J. Catal.* **110**, 58-73 (1988).
15. E. G. Derouane, Zeolites as Solid Solvents. *J. Mol. Catal. A: Chem.* **134**, 29-45 (1998).
16. J. M. Cowley, Scanning transmission electron microscopy of thin specimens. *Ultramicroscopy* **2**, 3-16 (1976)..
17. M. Haider *et al.*, Electron microscopy image enhanced. *Nature* **392**, 768-769 (1998).
18. R. Erni, M. D. Rossell, C. Kisielowski, U. Dahmen, Atomic-resolution imaging with a sub-50-pm electron probe. *Phys. Rev. Lett.* **102**, 96-101 (2009).
19. I. Lazić, E. G. T. Bosch, Analytical review of direct STEM imaging techniques for thin samples.

*Advances in Imaging and Electron Physics* **199**, 75-184 (2017).

20. Y. Zhu et al., Unravelling surface and interfacial structures of a metal-organic framework by transmission electron microscopy. *Nat. Mater.* **16**, 532-536 (2017).
21. D. Zhang et al., Atomic-resolution transmission electron microscopy of electron beam-sensitive crystalline materials. *Science* **359**, 675-679 (2018).
22. L. Liu et al., Imaging defects and their evolution in a metal–organic framework at sub-unit-cell resolution. *Nat. Chem.* **11**, 622-628 (2019).
23. X. Li et al., Direct imaging of tunable crystal surface structures of MOF MIL-101 using high-resolution electron microscopy. *J. Am. Chem. Soc.* **141**, 12021-12028 (2019).
24. Y. Li et al., Cryo-EM structures of atomic surfaces and host-guest chemistry in metal-organic frameworks. *Matter* **1**, 1–11 (2019).
25. H. Rose, Nonstandard imaging methods in electron microscopy. *Ultramicroscopy* **2**, 251-267 (1977).
26. I. Lazic, E. G. T. Bosch, S. Lazar, Phase contrast STEM for thin samples: Integrated differential phase contrast. *Ultramicroscopy* **160**, 265-280 (2016).
27. E. Yucelen, I. Lazic, E. G. T. Bosch, Phase contrast scanning transmission electron microscopy imaging of light and heavy atoms at the limit of contrast and resolution. *Sci. Rep.* **8**, 1-10 (2018).
28. E. M. Flanigen *et al.*, Silicalite, a new hydrophobic crystalline silica molecular sieve. *Nature* **271**, 512-516 (1978).
29. G. T. Kokotailo, S. L. Lawton, D. H. Olson, W. M. Meier, Structure of synthetic zeolite ZSM-5. *Nature* **272**, 437-438 (1978).
30. A. J. Jones, E. Iglesia, The Strength of Brønsted Acid Sites in Microporous Aluminosilicates. *ACS Catal.* **5**, 5741-5755 (2015) .
31. D. Lesthaeghe, V. Van Speybroeck, M. Waroquier, Theoretical Evaluation of Zeolite Confinement Effects on the Reactivity of bulky Intermediates. *Phys. Chem. Chem. Phys.* **11**, 5222-5226 (2009).
32. M. Boronat, A. Corma, What Is Measured When Measuring Acidity in Zeolites with Probe Molecules? *ACS Catal.* **9**, 1539-1548 (2019).
33. J. C. Taylor, Hexadeuteriobenzene locations in the cavities of ZSM-5 zeolite by powder neutron diffraction. *J. Chem. Soc. Chem. Commun.* **15**, 1186-1187 (1987).
34. R. Goyal, A. N. Fitch, H. Jobic, Powder neutron and X-ray diffraction studies of benzene adsorbed in zeolite ZSM-5. *J. Phys. Chem. B* **104**, 2878-2884 (2000).
35. L. Song, Z. L. Sun, H. Y. Ban, M. Dai, L. V. C. Rees, Studies of unusual adsorption and diffusion behaviour of benzene in silicalite-1. *Phys. Chem. Chem. Phys.* **6**, 4722-4731 (2004).
36. B. F. Mentzen, Localizing adsorption sites in zeolitic materials by X-ray powder diffraction: pyridine sorbed in B.ZSM-5. *J. Appl. Cryst.* **22**, 100-104 (1989).
37. S. Yuan *et al.*, Theoretical ONIOM2 study on pyridine adsorption in the channels and intersection of ZSM-5. *J. Phys. Chem. A* **109**, 2594-2601 (2005).

**Acknowledgments:**
This work was supported by the National Key Research and Development Program of China (2018YFB0604801 and 2018YFB0604803) and the National Natural Science Foundation of China (No.20141301065 and 21306103).